# A modeling approach to study the effect of cell polarization on keratinocyte migration

**Short title:** Modeling of keratinocyte migration

Matthias Jörg Fuhr[1*], Michael Meyer[2*], Eric Fehr[1], Gilles Ponzio[3,]

Sabine Werner[2**] and Hans Jürgen Herrmann[1**]

[1]ETH Zurich, Institute for Building Materials, Computational Physics for Engineering Materials, Schafmattstrasse 6, HIF E12, CH-8093 Zurich, Switzerland

[2]ETH Zurich, Institute of Molecular Health Sciences, Otto-Stern-Weg 7, HPL F12, 8093 Zürich, Switzerland

[3]CNRS, UMR 7275, 660, Route des Lucioles, Sophia Antipolis, 06560 Valbonne, France, and Université de Nice Sophia Antipolis, Nice, France

* Equal contribution

**Joint senior and corresponding authors

**Address for correspondence:**

Hans Jürgen Herrmann and Sabine Werner, Institute for Building Materials, Computational Physics for Engineering Materials (HJH) and Institute of Molecular Health Sciences (SW), ETH Zurich, 8093 Zurich, Switzerland.

E-mail Hans Jürgen Herrmann: hans@ifb.baug.ethz.ch

E-mail Sabine Werner: sabine.werner@biol.ethz.ch

**Abstract**

The skin forms an efficient barrier against the environment, and rapid cutaneous wound healing after injury is therefore essential. Healing of the uppermost layer of the skin, the epidermis, involves collective migration of keratinocytes, which requires coordinated polarization of the cells. To study this process, we developed a model that allows analysis of life-cell images of migrating keratinocytes in culture based on a small number of parameters, including the radius of the cells, their mass and their polarization. This computational approach allowed the analysis of cell migration at the front of the wound and a reliable identification and quantification of the impaired polarization and migration of keratinocytes from mice lacking fibroblast growth factors 1 and 2 – an established model of impaired healing. Therefore, our modeling approach is suitable for large-scale analysis of migration phenotypes of cells with specific genetic defects or upon treatment with different pharmacological agents.

**Introduction**

In response to skin injury, a complex wound healing process is initiated that aims to restore the injured body site. The repair process is divided into three partially overlapping phases, namely blood clotting and inflammation, new tissue formation, and finally tissue remodeling. Tissue formation includes the formation of a provisional dermal tissue, called granulation tissue, as well as coverage of the wound with a new epithelium (reepithelialization). The latter is initiated by migration of keratinocytes from the epidermis at the wound edge and from injured hair follicles, followed by proliferation of keratinocytes to replenish the missing tissue [1-3]. A defect in reepithelialization is characteristic for chronic human wounds, a frequent and severe pathology that mainly affects aged individuals as well as patients with diabetes, those treated with immunosuppressive compounds, or cancer patients treated with chemotherapy [4]. Interestingly, the reepithelialization defect in chronic ulcers is usually not due to impaired keratinocyte proliferation, but rather to a severe deficiency in the migratory response [5,6]. It is therefore highly relevant to characterize the effect of overexpression or loss of different genes or of external stimuli and pharmacological compounds on the migration process of these cells.

We recently described a novel mouse model for impaired wound healing, which – like the situation in chronic human wounds – is characterized by impaired keratinocyte migration in vitro and in vivo, but enhanced proliferation of these cells in vivo [7]. These mice lack fibroblast growth factor receptors (FGFR) 1 and 2 in keratinocytes [8], and therefore cannot respond to FGFs, which are key regulators of wound repair [9]. The migratory defect of FGFR1/2-deficient keratinocytes was also observed in cultured keratinocytes from these mice using scratch and transwell migration assays, while their proliferation rate in culture was not different compared to control cells [7,8]. A cellular and molecular characterization of the migrating cells identified defects in the formation of focal adhesions due to reduced expression of major focal adhesion components [8]. In addition, their polarization was impaired as reflected by fluorescence staining of the Golgi apparatus. While the Golgi was fully oriented towards the wound edge in 30% of the cells from control mice, this was only the case in 8% of the FGFR1/R2-deficient cells. This finding demonstrates a defect in cell polarization, although more subtle differences in polarization cannot be determined using this cell biological approach.

To further characterize and quantify the migration defect of FGFR-deficient cells and to analyze the migratory capacity of other cells, it is important to develop appropriate computational methods and models, taking into consideration the different behavior of cells within the cell monolayer ('interior cells') and those at the front ('border cells'). The cell locomotion of the interior cells is dominated by a coordinated 'flocking' movement, which depends on the cell density [10,11]. Simulations of self-driven particles, such as keratinocytes, suggest that short-range attractive-repulsive intercellular forces alone are sufficient to explain this coordinated movement [11]. Thereby, the cell motility undergoes a discontinuous kinetic phase transition from a disordered to an ordered state [12,13]. On the other hand, cells in the front row (border cells) can show both individual and collective behavior at the same time [14]. The individual behavior is characterized by cells, which dissociate from the scratch and individually explore the open space, while the collective one results in an explorative motion of the cell front. Simulations and experiments of Madin-Darby canine kidney cells showed that active 'leader' cells destabilize the scratch border by dragging their neighbors into the scratch, thereby facilitating closure [15,16]. However, the behavior of the border cells is still not completely understood. Therefore, it was the goal of the present study to establish a model that addresses this issue, and we focused on the analysis of the collective migration of border cells, since keratinocytes at the edge of skin wounds show this type of migratory behavior (Shaw and Martin, J Cell Sci 2009). Rather than segmenting single cells [17] or even signaling cascades [18], we decided to choose a particle model with each particle representing one cell. We present a reliable and reproducible model with few parameters that allows large-scale analysis of cells with different genetic defects or upon pharmacological manipulation.

**Materials and Methods**

**In vitro keratinocyte migration experiments**

Spontaneously immortalized keratinocytes from mice lacking FGFR1 and FGFR2 in this cell type had previously been published [8]. These established cell lines were used for migration studies. The mutant mice had been obtained by mating of mice with floxed *Fgfr1* and *Fgfr2* alleles with transgenic mice expressing Cre recombinase under control of the keratin 5 promoter (K5-Cre mice). Control cells were from mice with floxed *Fgfr1* and *Fgfr2* alleles, but lacking Cre recombinase. Therefore, the control cells express normal levels of both receptors. Maintenance of the mice as well as isolation of keratinocytes from these mice after sacrifice had been performed with permission from the local veterinary authorities (Kantonales Veterinärmat der Stadt Zürich). Keratinocytes were cultured in defined keratinocyte serum-free medium (dK-SFM) (Invitrogen, Carlsbad, CA) supplemented with 10 ng/ml EGF and $10^{-10}$M cholera toxin and penicillin/streptomycin (Sigma, Munich, Germany). For scratch wounding assays they were grown to confluency in the same medium. A scratch was made within the cell layer with a sterile pipette tip. For live cell imaging we used a Zeiss 200M microscope with a 10x 0.3NA Plan NeoFluar objective with an incubator box and a motorized stage. Micrographs were taken every 5 minutes. All migration experiments were performed under the same experimental conditions.

**Simulation of cell migration**

A molecular dynamic simulation was used to model the migration of keratinocytes. In order to limit the number of model parameters to a minimum, cells were modeled as spheres of fixed radius $R$ and mass $m$, having a short-range repulsive interaction of elastic type and a medium range attractive interaction with other cells. The short range potential simulates volumetric exclusion between cell bodies, whereas the medium range potential represents the cell-cell adhesion effects with subsequent inhibition of cell overgrowth due to contact inhibition. The cells move in continuous space following Newton's equation of motion

$$m \cdot \frac{d^2 x}{dt^2} = F_{Polarization} + F_{Interaction}.$$

The polarization and the interaction forces are given by

$$F_{Polarization} = P \cdot \vec{\zeta}$$

and

$$F_{Interaction} = 4\varepsilon \sum_{ij} \frac{\vec{r}}{r} \begin{cases} \dfrac{-\sigma}{r^2} & if \quad r < R \\ \dfrac{2\sigma^2}{r^3} & if \quad R \le r < r_c \end{cases},$$

where $P$ is the strength of the polarization given in J and does not change with time. $\vec{\zeta}$ is a unit vector with a direction chosen uniformly at random and given in 1/mm. $\varepsilon$ defines the scale of the cell-cell interactions and is given in J, $r = |\vec{r}| = |\vec{x}_i - \vec{x}_j|$ denotes the distance between the centers of the cells $i$ and $j$, and $r_c$ is the cut-off radius of the interactions. $\sigma$ is the zero crossing of the potential. This formulation of the interaction force does not include viscous properties in order to limit the model complexity. Initially, the cells were randomly placed on a regular grid on the left and right side of a square box in such a way that the wound in between had a width of approximately 8 − 10 cells, and the initial density in the in vitro experiments and in the simulation was similar. In total, the migration of 450 cells was simulated. This is a small number of cells compared to the about $10^7$ cells in the experiment, allowing easy simulation of the model. In fact, this limited number of cells corresponds approximately to the analyzed region. Moreover, this is the minimum number of cells required to describe a representative volume of the experiments. The initial direction and strength of the cell movement were chosen randomly. Periodic boundary conditions were imposed to all walls of the box. Newton's equations were solved explicitly using a Verlet scheme for time integration [19]. Simulation parameters are given in Table 1. The simulation was terminated upon wound closure.

**Measuring the cell monolayer-wound interface**

Digital image processing

Cell imaging was performed over a time period of 48 hours. Images with a size of 1024 x 1024 pixels were taken at regular time intervals. In order to measure the dynamics of cell migration, i.e. the behavior of the wound over time, we analyzed 60 time-lapse images, i.e. one image every 48 minutes, for each sample. In a first step, each time-lapse image was transformed into binary images, i.e. an image having only pixel values of "0" (air) and "1" (material), using a gray-level thresholding. Thereby, we determined for each pixel whether its gray-value is below (air) or above (cells) a specific threshold. A threshold of 0.5 was used. In the second step, the cell monolayer-wound interface (indicated in black in Fig. 1) was automatically determined by a morphological closing operation [20]. The resulting cell monolayer-wound interface is a line with a width of 1 pixel as indicated in black in Fig. 1.

Fractal dimension and roughness

The yardstick method [21] was used to determine the scaling behavior of the cell monolayer-wound interface, i.e. its fractal dimension $d_f$ and roughness $\varphi$. The fractal dimension is widely used in biology for the characterization of surfaces [Iannaccone and Khokha, 1996; Losa et al., 2005], and, therefore, the tissue-wound interface. For example, it has been used for wound healing assays of breast cancer cells [Sullivan et al., 2008]. The higher the fractal dimension, the higher the roughness and therefore the surface area of the cell monolayer-wound interface. Here, the roughness is defined as the exponent of the scaling law $RMS \propto S^{\varphi}$, where $RMS$ is the root mean square deviation of the cell monolayer-wound interface measured within a yardstick, i.e. a piece-wise linear approximation of the cell monolayer-wound interface by a line segment of size *S*. If the object is self-similar and the $RMS$ increases like a power-law for increasing *S*, an object is fractal.

Wound width

At t = 0 the left-hand side cell monolayer-wound interface is approximated by a line $Q$ using ordinary least squares. $Q$ can be represented either in the slope-intercept form or in vector form. The position of the left- and right-hand cell monolayer-wound interface at time t >= 0 is denoted by the scalars $q_L^{(t)}$ and $q_R^{(t)}$, respectively, given in µm and measured as the average distance of all the pixels of the cell monolayer-wound interface perpendicular to the line $Q$. Thus, the wound width $W^{(t)}$ is defined as the difference between $q_R^{(t)}$ and $q_L^{(t)}$, i.e. $W^{(t)} = q_R^{(t)} - q_L^{(t)}$.

**Results**

**Analysis of cell monolayer-wound interface**

We analyzed the migratory behavior of three independent keratinocyte cell lines from both control (ctr) and FGFR1/R2 knockout (ko) mice using in vitro scratch wounding assays. The migratory behavior of these cells had previously been characterized in detail [8]. The time evolution of the effective fractal dimension of the left-hand and right-hand cell monolayer-wound interface for various samples is shown in Fig. 2a and b, respectively. The best model fit was found using non-linear least squares. We observed that the fractal dimension $d_f$ increases in time for all ctr cell lines, whereas the increase is less pronounced for cells from ko mice. The right- and left-hand interfaces behave slightly differently for the ko cell lines. The fractal dimensions of the right-hand interfaces increase, whereas $d_f$ of the left-hand interfaces seem to stay constant. We assume that this difference is random and results from the limited number of samples that were analyzed (N=3). Furthermore, there is no biological evidence for a difference between both interfaces. Nevertheless, we were able to distinguish between ctr and ko cells by the slowed-down evolution of the effective fractal dimension of the latter.

**Analysis of wound closure**

For every segmented cell monolayer-wound interface we calculated its position in the direction perpendicular to the initial wound. In a first step we derived from this position of the right- and left-hand cell monolayer-wound interface the normalized wound width $\bar{W} = W^{(t)} / W^{(0)}$, the time evolution of which is presented for the various samples in Fig.

3. Again, as in the case of the effective fractal dimension, we observed that the wound closure of the ko cell lines is delayed compared to the ctr cell lines. While the wounds in the monolayer of ctr cells closed within about 20 hours, they stayed open during the entire experimental period (48 hours) in the monolayers of the ko cells. We were then able to quantify the dynamics by interpreting the movement of the cell monolayer-wound interfaces as a diffusive motion of keratinocytes. Hence, we defined an effective diffusion constant $D_{eff} = \dfrac{W^{(t)} \cdot \overline{l}}{2 \cdot t}$ per unit length $\overline{l} = 1$ μm, which is estimated by fitting a straight line to the measurement of $W^{(t)}$ in Fig. 3. The factor of 2 is included, because the wound is closed from both the left- and right-hand cell monolayer-wound interface. The diffusion is a measure of how fast the wound is closed. As shown in Table 2, we obtained effective diffusion constants for the ko cells that were approximately a factor of ten smaller than for the ctr cells.

**Simulation**

The time evolution of the cell simulation is shown in Fig. 1 for 0, 8 and 16 hours after scratching. 450 cells were initially placed in the box. Simulation parameters for the ctr cell lines are given in Table 1. The dynamics of the keratinocytes was simulated by choosing a different polarization P and by holding the depth of the potential $\varepsilon$ constant, i.e. $P/\varepsilon = 2$ and $P/\varepsilon = 1$ using a time step of 0.18 s and 0.432 s for the ctr and ko cells, respectively. These values were estimated on the basis of data from previous migration experiments (Meyer et al., 2012) and experience.

We observed that the evolution of the cell monolayer-wound interface of the simulations is consistent with the experimental data obtained in cell culture experiments using time-lapse microscopy. At the beginning of the experiments, the cell monolayer-wound interface is nearly straight and over time it forms a complex structure. This behavior is reflected in Fig. 2 by an increase in the fractal dimension with time. The evolution of the simulations' fractal dimension is perfectly consistent with the experimental data obtained by in vitro imaging. Moreover, the parameter $P$, i.e. the polarization, influences the cells such that a higher polarization leads to a higher mobility of the cells and a higher fractal dimension.

The time evolution of the wound closure is shown in Fig. 3a and b for the simulation of ctr and the ko cells, respectively. Consistent with the experimental data, the wound of the ctr cells was closed after approximately 20 hours of simulation, whereas the wound of the ko cells was still open. However, during the first 5 hours, there was a quantitative difference between the simulation and the microscopy analysis: In the cell simulation, the wound closure started immediately after injury, whereas in vitro only little cell movement was observed directly after scratching. This "reaction time", which is required for the initiation of cellular and molecular alterations required for migration, was not included in the cell simulation. Otherwise, all the following steps were consistent between the experimental and the modeling approach.

**Discussion**

We developed a novel computer simulation model to study keratinocyte migration in culture. Rather than mimicking all aspects of the system, the aim of our computer model was to reduce the complex biological system [17,18,24] into a simpler mathematical model, in order to investigate the effects of factors that are not isolated and directly adjustable in vitro. Here, we specifically analyzed the effect of cell polarization on migration, since polarization is important for the onset of migration as well as for completion of efficient wound healing. Since we used highly standardized conditions, the polarization of the cell was modeled by a single and static parameter, whereas the interaction between the cells was taken into account by a potential with both an attractive and a repulsive part. Since a static polarization is unlikely, this simple model allowed us to study the isolated effect of polarization on the development of the cell monolayer-wound interface and the dynamics of the wound closure.

By comparing simulations with the experimental data, we obtained two major results: (1) Cell polarization strongly affects both the structure of the cell monolayer-wound interface and the dynamics of wound closure in a keratinocyte monolayer and (2) increasing the polarization enhances the roughness of the cell monolayer-wound interface. The cell biological data allowed us to determine if cells are polarized or not polarized. Thus, compared to the experimental uncertainty, the sensitivity of the polarization parameter is negligible under our highly reproducible experimental conditions. In future studies performed using different conditions, e.g. different cellular density, coating with different matrices or supplementation of the medium with different growth factors, the parameter will have to be modified.

Considering the numerical data in detail, it is clear that $d_f(t=0)=1$ as the interfaces are created by straight scratches. Afterwards, the fractal dimension increases, because of the observed fingering instabilities. The normalized RMS deviation of the cell monolayer-wound interface is expected to scale as $RMS \propto S^\varphi$ as shown in the insets of Fig. 2. A roughness $\varphi=1$ and $\varphi \neq 1$ would correspond to a self-similar and a self-affine behavior, respectively. In fact, we observed two regimes. Considering finite-size scaling, the lower regime indicates $\varphi=1$ (solid lines in the inset). The upper regime and the location of the crossover result directly from the initial wound width, which is the characteristic length scale of the system and limits the fluctuations of both cell monolayer-wound interfaces.

The diffusion constant suggests that the cells from ko mice are less motile than cells from ctr mice. This is consistent with our previous experimental data showing impaired migration of these cells and a reduced migratory velocity [8]. As the underlying molecular mechanism we discovered a reduction in the expression of major focal adhesion proteins, resulting in impaired adhesion and migration [8]. Thus, our strategy allows for the evaluation of effects of changes in parameters already in silico, thereby reducing the number of required experiments. This will facilitate future studies that aim to address the mechanisms underlying various wound healing phenotypes by matching the model to the phenotype by adjusting the parameters so to fit the experimental findings. Given the importance of keratinocyte migration for efficient wound healing, this approach has obvious biomedical implications, since the consequences of alterations in gene function or of various external stimuli can be efficiently studied.

Despite this success, our model also has obvious limitations. In particular, the polarization itself cannot explain the dynamics of the cells directly after scratching. It is most likely that additional factors affect these early events in culture. These may include the remodeling of cell-cell and cell-matrix adhesions [25] and the secretion of extracellular matrix molecules and proteinases by keratinocytes [26], which allow the initiation of scratch wound healing [27]. Moreover, it is not possible to translate the observation of the polarization in the experiments directly to the model parameter. Therefore, future studies will focus on the development of a more realistic model, e.g. using a viscoelastic material for the whole cell such as the Voigt model or the homogenous standard linear solid model [28], including a time-dependency of the polarization. Including parameters for cell-cell adhesion and a variable for the presence of freshly secreted extracellular matrix would increase the quality of the model, but at the same time increase its complexity and limit the model`s capacity to study various sizes and shapes of wounds. On the other hand, expanding both the experiments and the model to the third dimension would significantly increase the relevance for in vivo wound healing.

**Acknowledgments**

We thank Troy Shinbrot for helpful discussion about cell models. This work was supported by grants from the Swiss National Foundation SNF (No. 205321-121701 to H.J.H. and 310030_132884 to S.W.) and the European Research Council (ERC) Advanced Grant 319968-FlowCCS (to H.J.H.).

**Table 1:** Model parameters for the simulation of cell migration of the ctr cells used throughout this work.

| Description | Symbol | Value | Unit |
|---|---|---|---|
| Radius | $R$ | 15*10^-6 | m |
| Mass | $m$ | 0.015*10^-9 | kg |
| Polarization | $P$ | $\frac{P}{\epsilon} = 2$ | J |
| Depth of the potential | $\epsilon$ | | J |
| Cut-off radius | $r_c$ | 60*10^-6 | m |
| Expected half-distance between the cell centers | $\sigma$ | 21*10^-6 | m |

**Table 2:** Effective diffusion constant for the left and right cell monolayer-wound interface using cell lines derived from ko or ctr mice.

| Genotype | Cell type | Effective diffusion constant [$\mu m^2/s$] | |
| --- | --- | --- | --- |
| | | Left boundary | Right boundary |
| FGFR1/R2 floxed | mKC ctr | 2.37 x 10 ^ -6 | 4.74 x 10 ^ -6 |
| FGFR1/R2 floxed | mKC ctr 1 | 3.38 x 10 ^ -6 | 6.37 x 10 ^ -6 |
| FGFR1/R2 floxed | mKC ctr 2 | 3.20 x 10 ^ -6 | 7.72 x 10 ^ -6 |
| K5Cre-FGFR1/R2 floxed | R1-R2-ko | 0.76 x 10 ^ -7 | 6.49 x 10 ^ -7 |
| K5Cre-FGFR1/R2 floxed | R1-R2-ko1 | 2.2 x 10 ^ -7 | 3.78 x 10 ^ -7 |
| K5Cre-FGFR1/R2 floxed | R1-R2-ko 2 | 4.1 x 10 ^ -7 | 5.41 x 10 ^ -7 |

**Figure Legends**

**Figure 1:** In vitro and in silico experiments to study the healing of scratch wounds in a keratinocyte monolayer. (a) Light microscopy image of a cell layer of keratinocytes initially wounded by a straight scratch and (b) molecular dynamic simulation. Colors mark different time steps of the cell front evolution into the wound, 0 (red), 8 (blue) and 16 (grey) hours after scratching.

**Figure 2:** Time evolution of the effective fractal dimension $d_f$ for (a) the left and (b) the right cell monolayer-wound interface. The structure of the cell monolayer-wound interface of the ko cells ($\square$,$\triangle$,$\mathbf{O}$, $d_f$ shifted by -0.1 for a better visibility) evolves significantly slower compared to the ctr cells (+,×, ∗). Simulations (filled symbols) agree with the experimental data. The solid and dashed lines fit the experimental and simulation data, respectively. The insets show the root mean square deviation (RMS) 12 hours after scratching, normalized by the initial wound width $W^{(0)}$ for both the experiment mKC ctr 2 ($\square$) and the corresponding simulation ($\mathbf{O}$). The solid lines are power laws with exponent 1, i.e. roughness $\varphi$ = 1. The size of the error bars represents the variability of the data within two standard deviations.

**Figure 3:** Time evolution of the normalized wound width $\bar{W} = W^{(t)} / W^{(0)}$ for (a) the ctr cells (●) and (b) the ko cells (○), where the red symbols represent simulations.

**Figure 1**

Fig. 1

(a)

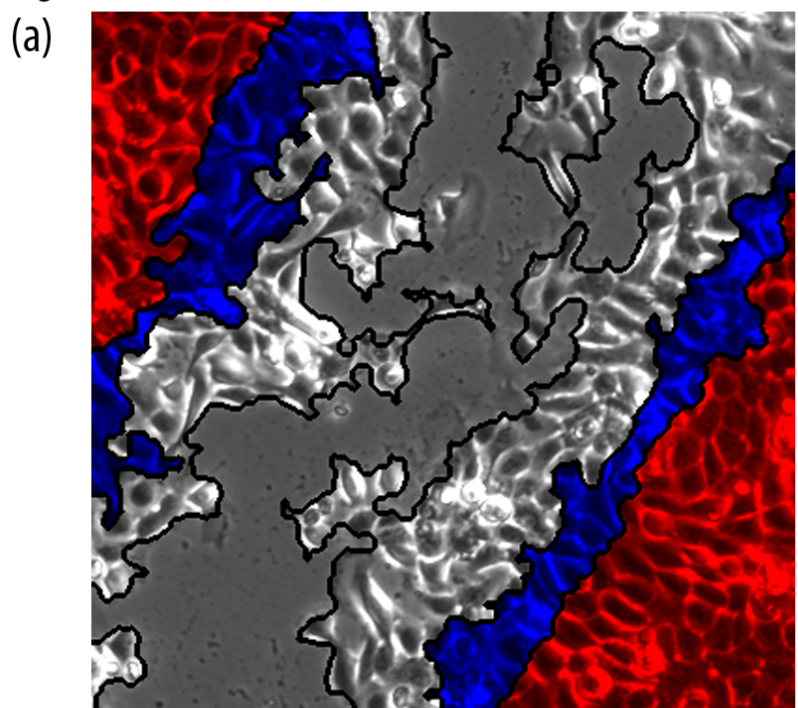

(b)

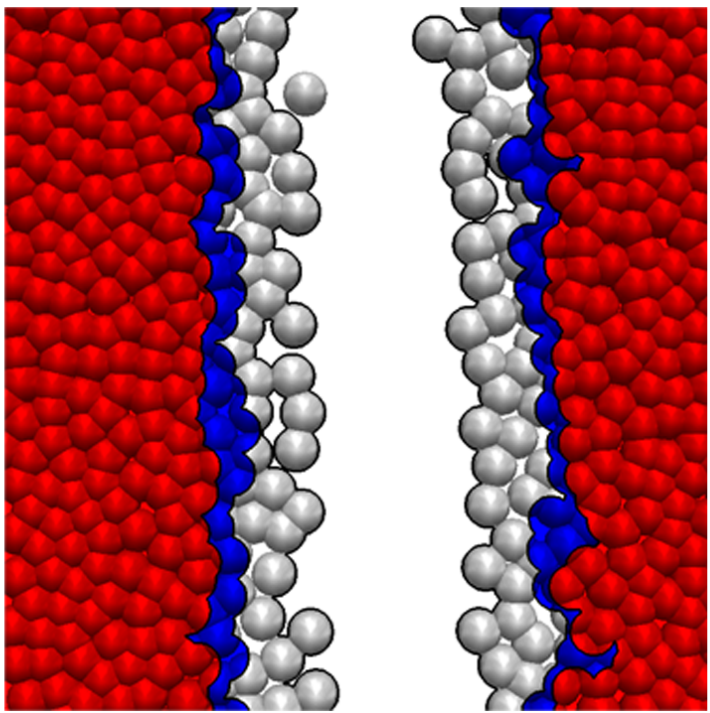

**Figure 2a**

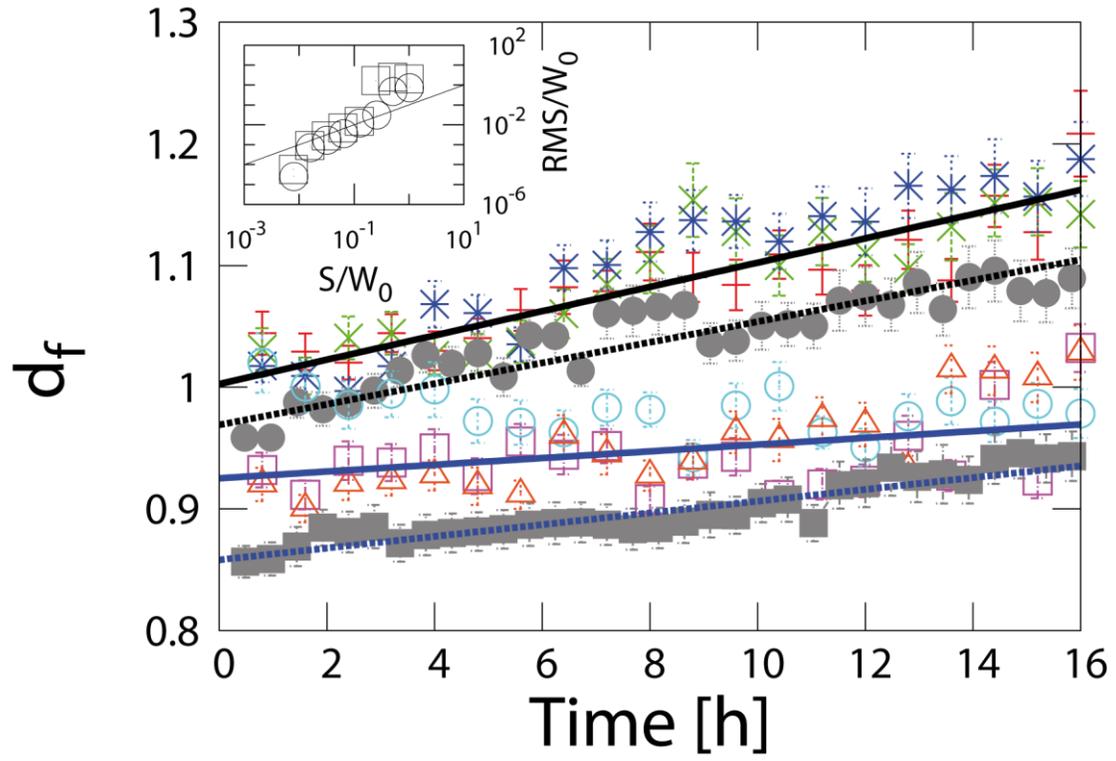

**Figure 2b**

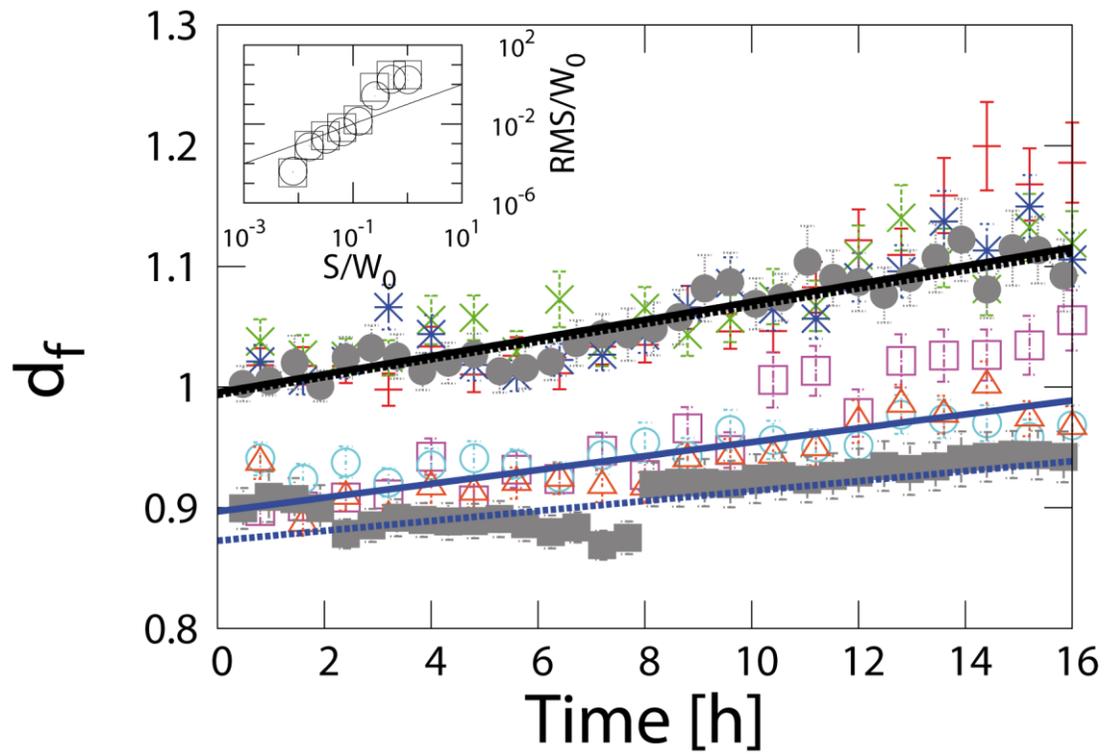

**Figure 3**

Fig. 3

(a)

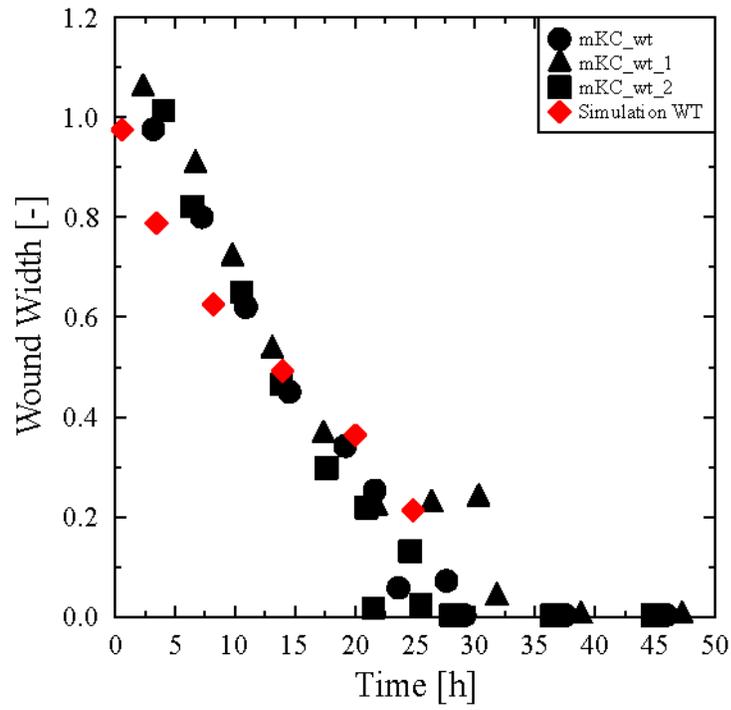

(b)

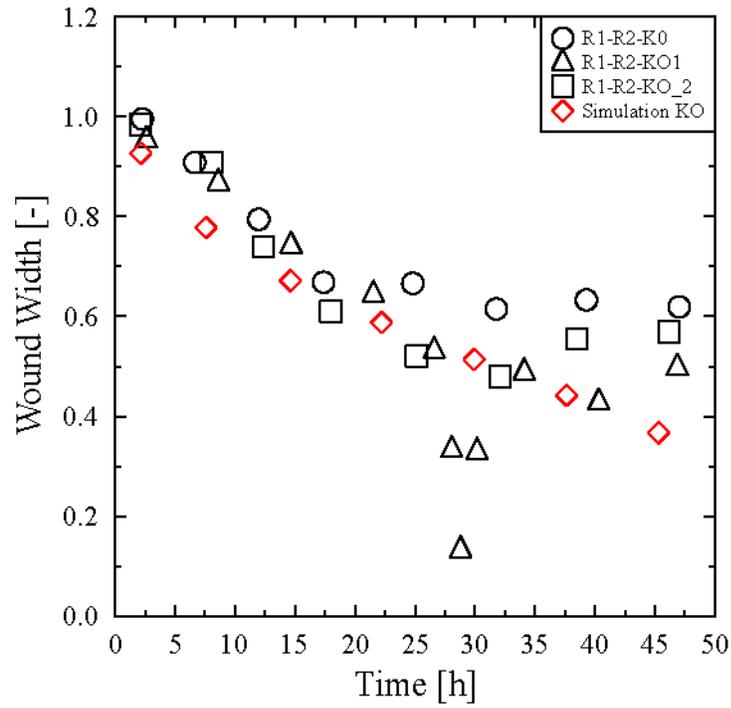